\documentclass[12pt,a4paper,reqno]{article}
\usepackage{amsmath}
\usepackage{amssymb}
\usepackage{amsthm}
\usepackage{graphicx}

\newcommand{\upline}{\vspace{-\abovedisplayskip}\vspace{-\baselineskip}}
\renewcommand{\(}{\left(}
\renewcommand{\)}{\right)}
\newcommand{\pr}{^\prime}
\newcommand{\e}{\mbox{e}}
\renewcommand{\hbar}{\hslash}

\renewcommand{\S}{\mathcal{S}}
\newcommand{\tR}{\text{R}}
\newcommand{\tB}{\text{B}}
\newcommand{\ta}{\text{a}}
\newcommand{\tf}{\text{f}}
\newcommand{\br}{\mathbf{r}}
\newcommand{\bE}{\mathbf{E}}
\newcommand{\ox}{\otimes}
\newcommand{\<}{\langle}
\renewcommand{\>}{\rangle}

\renewcommand{\Re}{\operatorname{Re}}
\renewcommand{\Im}{\operatorname{Im}}
\newcommand{\tr}{\operatorname{tr}}

\theoremstyle{plain} 

\theoremstyle{definition}

\theoremstyle{remark}

\begin{document}
\title{Diese Verdammte Quantenspringerei}
\author{Anthony Sudbery\\[10pt] \small Department of Mathematics,
    University of York,\\[-2pt] \small Heslington, York, England YO10 5DD\\ 
    \small  Email:  as2@york.ac.uk}
\date{23 October 2000}

\maketitle

\begin{abstract} It is argued that the conventional formulation of
quantum mechanics is inadequate: the usual interpretation of the
mathematical formalism in terms of the results of measurements cannot be
applied to situations in which discontinuous transitions (``quantum
jumps") are observed as they happen, since nothing that can be called a
measurement happens at the moment of observation. Attempts to force such
observations into the standard mould lead to absurd results: ``a watched
pot never boils". Experiments show both that this result is correct when
the experiment does indeed consist of a series of measurements, and that
it is not when the experiment consists of a period of observation:
quantum jumps do happen. The possibilities for improving the formulation
by incorporating transitions in the basic postulates are reviewed, and a
satisfactory postulate is obtained by modifying a suggestion of Bell's.
This requires a distinction between the external description of the
whole of a physical system and internal descriptions which are
themselves physical events in the system. It is shown that this gives
correct results for simple unstable systems and for the quantum-jump
experiments. 
\end{abstract}

\newpage

\section{Introduction}

In 1929 Erwin Schr\"odinger complained ``If I had known we were going 
to go on having all this damned quantum-jumping, I would never have got 
involved in the subject'' \cite{Schrquote}. Over twenty years later he was 
still not reconciled to the idea, posing the question ``Are there quantum 
jumps?'' as the title of a paper in the {\em British Journal for the Philosophy 
of Science}. After another 34 years the question was definitively answered 
in another title when Nagourney, Sandberg and Dehmelt published their 
paper ``Shelved optical electron amplifier: observation of quantum jumps" 
in {\em Physical Review Letters}. But in spite of the direct and visible 
evidence for the reality of quantum jumps provided by this beautiful experiment, 
their place in theory is still not clear. The mathematical formulation 
of quantum mechanics is just as continuous as Schr\"odinger could have 
wished; the jumps, if they are present in the theory at all, are there 
only as part of the interpretive framework. In this paper I want to examine 
the necessity for and the possibility of an element of discontinuity in 
the theoretical framework of quantum mechanics. 

To a particle physicist one of the most paradoxical aspects of quantum 
mechanics is that it is the only theoretical framework for one's subject
 and yet it does not acknowledge the fundamental empirical elements in 
which one is interested. The evidence which must be explained by
theories of  elementary particle physics consists of {\em events} like
the decay of  the $\Omega^-$ particle captured in the bubble chamber
photograph of Figure  1. But {\em there are no events in quantum
theory}.\footnote{The word ``event" sometimes occurs in the sense of
probability theory, with the meaning of ``a statement being true" -- for
example, van Fraassen \cite{vanFraassen} claims to base his modal
interpretation on events, but defines an event as the truth of a
statement of the form ``Observable $B$ has value $b$". Throughout this
paper I will use ``event" with its primary English meaning of
``happening".} The nearest thing to an event described in the basic
theory  is the result of a measurement --- something which has been
deliberately  provoked by an experimenter's conscious action. There is
no way that the  theory can describe events, like the $\Omega^-$ decay
in Figure 1, which  happen spontaneously and are passively recorded by
the waiting experimenter  --- which is how the photograph of Figure 1
was obtained. 

\begin{figure}
\begin{center}
  \includegraphics[width=10cm]{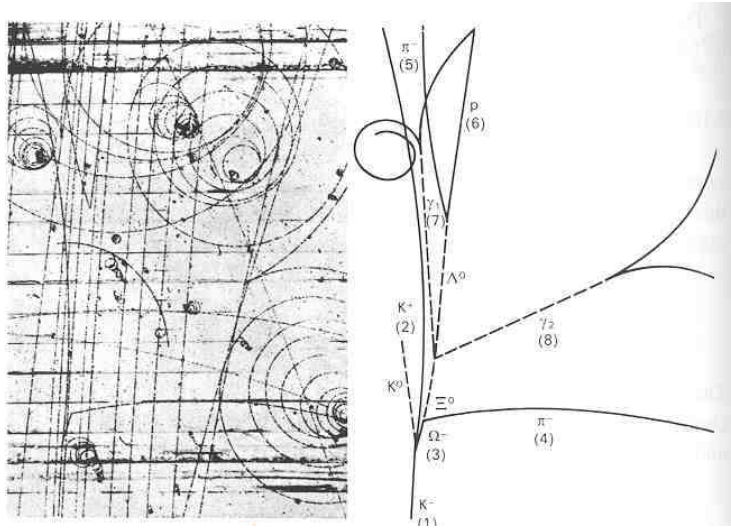}
  \caption{The world's first $\Omega^-$ event (Brookhaven National
    Laboratory, 1964).}
\end{center}
\end{figure}

Nevertheless, textbooks do purport to derive decay rates for such events 
from the general principles of quantum mechanics. The derivation goes 
like this. To describe a decay $A \rightarrow B+C$ we start at time $t=0$ 
with a state $|A\> $ in which the unstable particle $A$ is certain 
not to have decayed, and follow its time evolution, governed by a Hamiltonian 
$H$, to a superposition of the initial state and a state of the decay 
products $B$ and $C$:
\begin{equation}\label{decay1}
\e^{-iHt}|A\> = a(t)|A\> + b(t)|BC\> .
\end{equation}
The general principles of quantum mechanics are then supposed to yield 
the interpretation that $|b(t)|^2$ is the probability that by time $t$ 
there has been a  transition from particle $A$ to particles $B+C$. 

But where does this notion of a ``transition'' come from? It appears nowhere 
in the general principles of the theory, as usually stated. The literal 
application of these principles to the state \eqref{decay1} gives only the 
statement that {\em if a measurement is made at time $t$} (to determine, 
say, whether a particle of type $B$ is present), then $|b(t)|^2$ is the 
probability that the result of the measurement will be positive. The inference 
that something discontinuous (a transition) happened between time 0 and 
time $t$ is completely unwarranted. According to the official principles, 
quantum systems evolve continuously (as the time dependence of \eqref{decay1} 
shows), and quantum jumps occur only when provoked by the intervention 
of an experimenter. Despite Schr\"odinger's famous satire in his story 
of the cat, this 
remains the only description that the conventional form of the theory 
can tolerate; the idea of a transition must be a deplorable backsliding 
into classical habits of thought.

If this is what official quantum mechanics pronounces, then so much the 
worse for official quantum mechanics. In the actual physical situation that 
we are trying to describe, it is talk of ``a measurement at time $t$'' 
that is unwarranted. If a measurement can be said to take place at all 
when a decay is observed, it is an extended process occupying an interval 
of time rather than an instant; and the empirically meaningful time $t$ 
is not the time at which the experimenter decides to make a measurement 
on the system, but the time at which the system does something for the 
experimenter to observe.

Of course, there is no questioning the success of the theory in predicting 
the right transition rates. The procedure that starts with the time-dependent 
state vector \eqref{decay1}, produces from it a time-dependent probability 
$P(t)$, interprets this as a probability of something {\em having happened}, 
and on the strength of this interpretation derives from $P(t)$ a transition 
probability per unit time, certainly ends up with an empirically adequate 
result. The problem I want to discuss is that of formulating the theory 
so as to make this argument as sound as its conclusion. I will argue that
the germ of a solution was provided by John Bell in 1984
(\cite{Bell:beables}; see also \cite{QMPN} and \cite{Bub:book}), but
that it needs to be transposed into a different interpretative framework
in order to be fully satisfactory. In this paper I will do no more than suggest
how this transposition might be achieved.

Before discussing Bell's formulation, however, I would like to sharpen 
the problem by reviewing (a) the result of taking the conventional interpretation 
literally and pushing it to what seems to be its logical conclusion, and 
(b) the experiments which demonstrate the reality of quantum jumps.

\section{Quantum Jumps and Watched Pots: Theory and Experiment}

\medskip
\begin{center} A WATCHED POT NEVER BOILS \end{center}
\smallskip

Let us imagine ourselves in the situation of steadily watching the unstable 
particle $A$ in the process $A \rightarrow B+C$ --- suppose it is a radioactive 
nucleus surrounded by a spherical detector wired to a loudspeaker, and 
we are listening for the click which will announce that the decay has 
occurred. Being attentive observers, we know at every instant of our watch 
whether we have heard a click or not. How can this situation be described 
in conventional quantum mechanics? It seems to require that a measurement 
is made at every instant, our knowledge at each instant being the result 
of the corresponding measurement. But this involves an uncountable number 
of measurements in a finite interval, which, as well as being embarrassing 
to analyse, probably flatters our vigilance. Let us compromise by considering 
a finite sequence of measurements and trying to capture the ideal of continuous 
awareness by letting the measurements get closer and closer in time, taking 
the limit as the separation between them tends to zero. We will replace 
steady watching with repeated looking.

The result is disconcerting \cite{Yourgrau, MisraSuda, obsdecay}. We suppose that in the 
time interval $[0,T]$ there are $N+1$ measurements at times $0=t_0, t_1, 
\ldots, t_N=T$.  Let $p_n$ be the probability that the measurement at time 
$t_n$ gives the result ``$A$'' (of the two possible results ``$A$'' or 
``$B+C$''), i.e. the probability that at this time we are aware that 
no decay has occurred. According to the projection postulate of conventional 
quantum mechanics, the measurement causes the system to jump into one 
of the states $A$ or $B+C$, depending on the result of the measurement, 
after which it resumes (or rather --- and this is crucial --- {\em restarts}) 
its evolution following the Schr\"odinger equation. As a consequence of the 
unitary evolution in quantum mechanics, the resulting probabilities
always satisfy:

{\em Theorem} \cite{MisraSuda, obsdecay} In the limit as the dissection
$\{t_0, \ldots , t_N\}$ becomes infinitely fine, $p_N - p_0 \rightarrow 0$.

So if the unstable particle is observed assiduously enough, the probability 
that it has decayed at time $T$ is the same as at time $0$, namely zero 
if we start out knowing that it has not decayed. A watched pot never boils. 
Either that, or continu{\em al} measurement is not a good model for continu{\em 
ous} observation.

\medskip
\begin{center} WATCHED POTS IN THE LABORATORY\footnote{For a good recent
review see \cite{Whitaker:PQE}.} \end{center}
\smallskip
The conclusion of the above theorem has come to be known as the quantum 
Zeno paradox --- an unfortunate name, since Zeno's paradox of the arrow 
was resolved by Newton in classical mechanics and does not even arise 
in quantum mechanics. (Zeno's argument was that if one only considers 
the configuration of a physical object in space, there is no difference 
between a stationary arrow and a moving one and therefore no reason why 
any arrow should behave differently from a stationary one. Newton's answer 
is that position is not enough: at any instant, velocity is an independent 
property of an object which must also be specified before the equations 
of motion can be solved, these being second-order differential equations. 
In quantum mechanics, this does not apply. 
The equation of motion is first-order and a specification 
of configuration (in the form of a wave function $\psi (x)$) {\em is} enough. 
So Zeno's paradox is purely classical, in physical terms as well as historical 
ones.)

The proverbial wisdom that a watched pot never boils, on the other 
hand, becomes an objective fact only in quantum mechanics. This is {\em 
not} because of the role played by measurement --- the significance of 
measurement in the proof of the watched-pot theorem is the change in the 
probabilities brought about by a change in knowledge, and this is as much 
a feature of classical probability as of quantum mechanics. The distinctive 
contribution of quantum mechanics to the calculation is the form of the 
transition probability $P(t)$ as a function of time. Because it is the 
square of the inner product between two vectors which are initially orthogonal, 
it behaves like $t^2$ for small $t$ and is therefore slower to get started 
than a classical Poisson process, which behaves like $t$; if it is continually 
forced to reset while in this vulnerable early stage of its life, it never 
gets off the ground at all.

Experiments can test this distinctive small-time behaviour and
demonstrate the watched-pot phenomenon which can be uncontroversially
deduced from it, without any appeal to dubious projection postulates,
in a situation where there clearly are very short, rapidly repeated
physical processes which can be treated as measurements. The
experiments have been performed by Itano et al. \cite{watched-pot:expt} and
confirm the theorem: the effect of very rapidly repeated, very
short-duration interventions on an evolving quantum system is to freeze
the evolution. This is the first arm of the disjunction at the end of
the previous section. In the same and other laboratories the second arm of the
disjunction has also been confirmed: if the experimental situation is
one of passive, low-energy observation rather than active, high-enery
probing, then the evolution is not affected, there is no watched-pot
effect, and quantum jumps do occur. 

\begin{figure}
\unitlength=1.00mm
\special{em:linewidth 0.4pt}
\linethickness{0.4pt}
\begin{picture}(125.00,80.00)(0,50)
\put(50.00,60.00){\line(1,0){40.00}}
\put(60.00,65.00){\line(-2,3){20.00}}
\put(42.00,96.00){\line(2,-3){20.00}}
\put(20.00,100.00){\line(1,0){40.00}}
\put(60.00,100.00){\line(0,0){0.00}}
\put(80.00,130.00){\line(1,0){40.00}}
\put(77.00,66.00){\line(2,5){23.67}}
\put(75.00,68.00){\line(0,-1){6.00}}
\put(75.00,62.00){\line(5,3){5.00}}
\put(38.00,90.00){\line(1,6){1.33}}
\put(39.33,99.00){\line(2,-1){7.67}}
\put(55.00,65.00){\line(3,-1){9.00}}
\put(64.00,62.00){\line(-1,6){1.33}}
\put(125.00,130.00){\makebox(0,0)[lc]{$|\psi_3\>$}}
\put(95.00,60.00){\makebox(0,0)[lc]{$|\psi_1\>$}}
\put(15.00,100.00){\makebox(0,0)[rc]{$|\psi_2\>$}}
\put(98.00,125.00){\line(4,3){4.00}}
\put(102.00,128.00){\line(1,-5){1.00}}
\end{picture}
\caption{The atomic levels used in the quantum-jump experiment}
\end{figure}
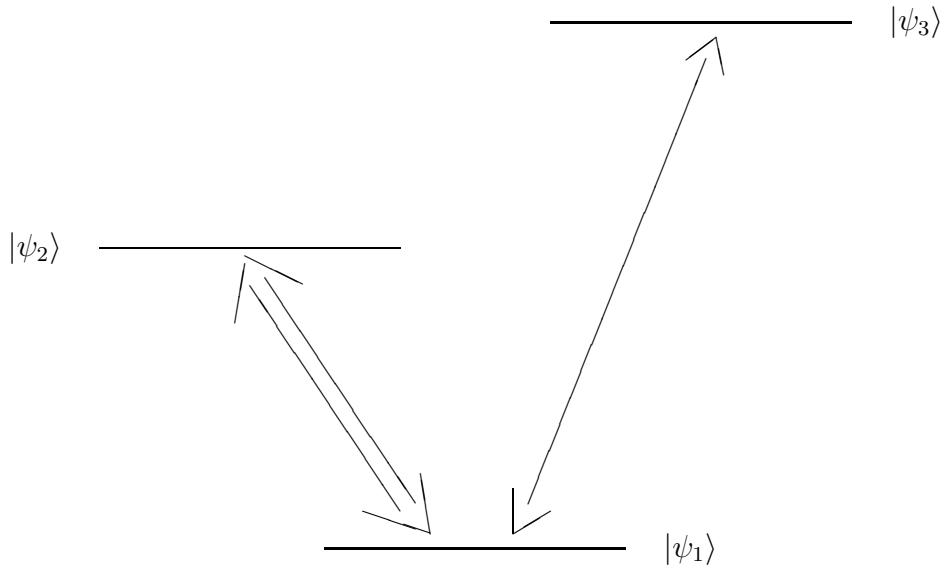

In the quantum-jump experiment \cite{qjumps:expt1, qjumps:expt2, qjumps:expt3},
originally proposed in 1975 by Dehmelt \cite{Dehmelt:qjumps} as a
practical method of detecting weak atomic transitions, a
single trapped ion is observed as it makes transitions between three
energy levels. The relevant levels, illustrated in Fig.\ 2, are a ground
state and two excited states, one strongly coupled and one weakly
coupled to the ground state. Let $E_0$, $E_1$ and $E_2$ be the energies
of the ground state, the strongly coupled excited state and the weakly
coupled excited state, and let us suppose that
the frequencies $(E_1 - E_0)/h$ and $(E_2 - E_0)/h$ are those of red and
blue light respectively. The ion is illuminated by a laser beam tuned to
the red frequency; absorption and stimulated emission of the laser mode
cause the ion to oscillate between the ground state and the first
excited state. Since these states are strongly coupled, the excited
state also spontaneously emits photons at the red frequency but in
other directions than that of the laser beam; these photons, which are
emitted at a rate of thousands per second, enable the experimenter to
see directly that the ion is in the subspace spanned by the ground state and
the first excited state. 

The ion is now illuminated by a second light source, low in intensity
and tuned to the blue frequency, which can stimulate transitions from
the ground state to the second excited state (the ``shelf" state). Since
this state is weakly coupled to the ground state, it has a long lifetime
and the ion will stay on the shelf for some time. This is observed as an
interruption in the red fluorescence from the first excited state.
Seeing the sudden start of this dark period is an observation of a
quantum jump from the ground state to the shelf; its equally sharp end
marks a jump from the shelf down to the ground state. 

The watched-pot experiment \cite{watched-pot:expt} uses a similar ion,
but the roles of the two states are reversed. The continuous laser beam
is now tuned to the blue frequency, so that the ion oscillates between
the ground state and the weakly coupled second excited state; the
weakness of the coupling makes the oscillation slow. This oscillation is
regarded as the natural evolution of the two-state system formed by the
ground state and the second excited state of the ion. The experiment
consists of measurements to determine which of the two states the system
is in, using intense laser pulses of light at the red frequency. If the
ion is in the ground state, it will absorb a photon from the red pulse
and be excited to the first excited state. Observation immediately
afterwards of an emitted red photon, not in the direction of the laser
beam, provides the result of the measurement (``ground state") and, in
accord with the projection postulate, leaves the system in the ground
state. On the other hand, if the ion was in the second excited state
then the red pulse will have no effect on it. Observation of no red
photon provides the result ``second excited state" and leaves the system
in that state. 

In Itano and Wineland's experiment a large number of such ions,
initially in the ground state, were illuminated by the blue laser and
allowed to oscillate without measurements for half a period. By
switching off the blue laser beam and counting the photons emitted in
the subsequent (slow) decay to the ground state, it was confirmed that
the ions had all evolved as expected and reached the second excited
state. Then the same evolution was subjected to repeated measurements by
means of a large number of short red laser pulses. At the end of the
same half-period, it was found that all of the ions were now in the
ground state. The watched pot had not boiled.

\medskip
\begin{center} WATCHED POTS IN THE STUDY \end{center}
\smallskip

What is the theoretical difference between these two experiments, and 
why doesn't the watched-pot theorem apply equally to both? In both cases 
we have an object system which is under investigation and whose state 
would, if left to itself, develop according to the Schr\"odinger equation:
\[
|\psi_0\> \rightarrow |\psi (t)\> = a(t)|\psi_1\> + b(t)|\psi_2\>
\] 
where $|\psi_1\> $ and $|\psi_2\> $ are experimentally distinguishable 
states. We need to go deeper than the superficial interpretation which 
declares that a measurement at time $t$ will have a result (1 or 2) and 
that after the measurement the system will be in the appropriate state 
$|\psi_1\> $ or $|\psi_2\> $. Following von Neumann, we must include 
the apparatus in the quantum description and consider a combined system 
of the object together with the apparatus, with product states $|\psi\> 
|\alpha \> $ where $|\alpha \> $ denotes a state of the apparatus. A 
measurement is brought about by an interaction between the object and 
the apparatus which changes the combined state as follows:
\begin{eqnarray*}
|\psi _1\> |\alpha _0\> \rightarrow |\psi _1\> |\alpha 
_1\> \\
|\psi _2\> |\alpha _0\> \rightarrow |\psi _2\> |\alpha 
_2\> \end{eqnarray*}
where $|\alpha _0\> $ is the initial (``ready'') state of the apparatus 
and $|\alpha _1\> $ and $|\alpha _2\> $ are the states in which 
it registers the results 1 and 2. Thus the whole process of making a measurement 
on a changing object system can be represented as 
\[ 
|\psi_0\> |\alpha_0\> \rightarrow |\psi(t)\> |\alpha_0\> 
\rightarrow a(t)|\psi_1\> |\alpha_1\>  + b(t)|\psi_2\> |\alpha_2\> . 
\]

It is clear from this that the Hamiltonian for the interaction 
between the apparatus and the object must be time-dependent; the apparatus 
has to be switched on at time $t$, when the measurement starts. Here is 
the difference between the watched-pot experiment and the quantum-jump 
experiment. In the first the interaction between the object and the apparatus 
is indeed time-dependent; the apparatus consists of the pulsed laser beam, 
and there is time dependence in the pulsing. In the quantum-jump experiment 
there is no time dependence: there is a steady laser beam, and a photon 
detector which is switched on before the experiment begins and is in a 
constant state of readiness to respond. Thus the theoretical measurement 
events at times $t_0, \ldots , t_N$ in the watched-pot theorem fairly 
describe the physical reality in the watched-pot experiment, but they 
have no place in the quantum-jump experiment and it should be no surprise 
that the conclusion of the theorem is not verified. 

A theoretical analysis \cite{obsdecay} of continuous observation  of an
unstable system, with a constant interaction between the observing 
apparatus and the decaying system, gives some reason to believe in the
watched-pot  effect even in this situation. The crucial parameter is the
response time  of the observing apparatus in comparison to the time
constants of the  decaying system. It is not the lifetime $T$ of the
decay that is important, but the time $\tau$ during which the decay
probability behaves quadratically, before the onset of the exponential
regime which is all that is usually observed. If the  apparatus responds
much faster than $\tau$, then its mere presence is enough to inhibit the
decay. However, since  the non-exponential decay time $\tau$ is several
orders of magnitude smaller  than the lifetime and has never been
observed in real unstable systems,  it is difficult to test this version
of the watched-pot effect.

\section{Transition Postulates}

\medskip
\begin{center} NECESSARY TRANSITION PROBABILITIES \end{center}
\smallskip

 We are left with a theoretical problem. If we cannot use the notion of 
``measurement at time $t$'' to explain the meaning of a time-dependent 
state vector, what can we say that will justify or codify what physicists 
do with such vectors, and that will predict our experience of quantum 
jumps? A simple answer is that transitions must be given a fundamental 
role in the theory; one of its basic postulates should be of the form 
``If the system [however broadly conceived] is in state $\psi$ at time 
$t$, there is a probability $T_{\phi \psi}(t)dt$ that it will make a transition to 
state $\phi$ between $t$ and $t+dt$''. Such a postulate, if it is to be 
fundamental, would need to be accompanied by a clear statement of exactly 
what the eligible states $\phi, \psi$ are. We certainly need something 
less {\em ad hoc} than the prescription students are left to deduce from 
current textbook paradigms, in which it is tacitly assumed that systems make 
transitions between eigenstates of $H_0$, a reference Hamiltonian which 
is chosen for no more principled reason than its convenience or familiarity.

A suggestion of this kind was made by John Bell in 1984 \cite{Bell:beables}.
I will present it in a generalised form \cite{QMPN, determlimit,
Bub:book}. The basic idea is that there is a set of special physical
quantities, which have a fundamental status; Bell liked to call them
\emph{beables} (as opposed to ``observables'') to emphasise their
objective nature (though the proposal is not just that they can be but
that they do be). These quantities always have definite
values, which change stochastically according to transition
probabilities which are determined by the solution to the Schr\"odinger
equation. Equivalently, one can replace the special quantities by a special set of
subspaces of state space, namely their eigenspaces, and the actual values of the
special quantities by a vector (the projection of the full state vector)
in the corresponding subspace. I will call these subspaces the
\emph{viable} subspaces. They may vary with time.

Then the complete description of a physical system at any
instant of time has two parts: a state vector in one of the viable subspaces, 
which is what we are aware of and which I will call the {\em visible} 
state; and the solution of the Schr\"odinger equation, in general a superposition 
of possible visible states, which guides the transition of the visible
state between the different subspaces. I will call the latter the {\em pilot} state
(following Bell \cite{Bell:pilot}, who used this term to describe the role of the wave
function in the de Broglie/Bohm theory, a special case of this theory in which 
the visible states are eigenstates of position \cite{determlimit}).

\medskip
\begin{center} POSSIBLE TRANSITION PROBABILITIES \end{center}
\smallskip

We know how the pilot state changes in time: it satisfies the
Schr\"odinger equation. What are the rules governing the change of the
visible state? We will now, following Bacciagaluppi and Dickson
\cite{BacciaDickson}, discuss the possible transition probabilities
which are consistent with the usual quantum-mechanical rules for the
results of measurements. 

Let $|\Psi(t)\>$ be the pilot state, let $\S_1(t), \S_2(t),
\ldots$ be the viable subspaces at time $t$, and let 
$\Pi_m(t)\>$ be the projection onto $\S_m$. Then the visible state at
time $t$ is one of the states $|\psi_m(t)\>=\Pi_m|\Psi(t)\>$,
and there is a set of positive real numbers $T_{nm}$ such that for 
$m \ne n$, the probability of transition from $|\psi_m(t)\>$ at time $t$
to $|\psi_n(t+\delta t)\>$ at time $t + \delta t$ is $T_{nm}\delta t$
(i.e. $T_{nm}\delta t$ is the probability that the visible state is
$|\psi_n(t+\delta (t)\>$ at time $t + \delta t$ if it is $|\psi_m(t)\>$ 
at time $t$). Let $P_m(t)$ be the probability that the visible state is
$|\psi(t)\>$ at time $t$; this then satisfies the master equation
\begin{equation}\label{master}
\frac{dP_m}{dt} = \sum_n(w_{mn} - w_{nm})
\end{equation}
where $w_{mn} = T_{mn}P_n$.

We would like to identify $P_m(t)$ with the quantum-mechanical probability
\begin{equation}\label{qmprob}
P_m(t) = |\<\Psi(t)|\psi_m(t)\>|^2
= \left|\<\Psi(t)|\Pi_m(t)|\Psi(t)\>\right|^2
\end{equation}
which gives
\[
\frac{dP_m}{dt} = 2\Re \left[\<\Psi(t)|\Pi_m(t)|\frac{d\Psi}{dt}\> +
\<\Psi(t)|\frac{d\Pi_m}{dt}|\Psi(t)\>\right]
\]
Since the projectors $\Pi_m$ are complete, this can be written as 
\[
\frac{dP_m}{dt} = \sum_n J_{mn}
\]
where
\begin{equation}\label{trans1}
J_{mn} = 2\Im \<\Psi|\(\hbar^{-1}\Pi_mH\Pi_n +
\frac{d\Pi_m}{dt}\Pi_n\)|\Psi\>,
\end{equation}
$H$ being the Hamiltonian. Then the orthogonality $\Pi_m\Pi_n = \delta_{mn}$ gives
\[
J_{mn} = -J_{nm}
\]
so it is possible to find $w_{mn}$ such that 
\[
w_{mn} - w_{nm} = J_{mn} \qquad \text{and} \qquad w_{mn} \ge 0.
\]
One possible solution is
\begin{equation}\label{trans2}
w_{mn} = \max (J_{mn}, 0)
\end{equation}
but any positive symmetric $x_{mn}$ could be added to this. More
generally, $x_{mn}$ need not be symmetric but must satisfy
$\sum_n(x_{mn} - x_{nm}) = 0$.

The transition probabilities \eqref{trans2} were originally suggested by
Bell \cite{Bell:beables} and extended to
the time-dependent case by Bacciagaluppi and Dickson
\cite{BacciaDickson}. Vink \cite{Vink}
has investigated more general possibilities and shown that they include
Nelson's stochastic dynamics \cite{Nelson:book}. It has been shown 
\cite{determlimit} that the Bell transition probabilities \eqref{trans2}
yield the deterministic de Broglie-Bohm theory in the continuum limit of
a lattice model of a particle moving in a potential. Bell's solution has
the feature that between any given pair of states, the transitions are
all one-way; if there can be a transition from $|\psi_m\>$ to
$|\psi_n\>$, there cannot be a transition from $|\psi_n\>$ to
$|\psi_m\>$. As we will now see, this is a natural feature of decay
processes.

\medskip
\begin{center}APPLICATION TO DECAY PROCESSES\end{center}
\smallskip

 Let us return to the problem of the unstable particle and model the
decay  $A\rightarrow B+C$ a little more realistically. If the decay
products  had just the one state $|BC\>$, the hermiticity of the
hamiltonian  would guarantee that the initial state $|A\>$ would be
regenerated.  For an irreversible decay of the kind we observe, the
decay products  must have an infinite-dimensional state space
$\S^\prime$ in which they can disperse \cite{rapid-dispersal}. This
makes it possible to suppose that the time of decay can be measured by
examining the decay products.  We can continue to suppose that the
unstable particle $A$ has just one  state, coupled by the decay
Hamiltonian to a decay state $|BC(0)\>\in \S^\prime$  which then evolves
by the internal dynamics of the $BC$ system to $|BC(t)\>\in \S^\prime$
after time $t$ (as well as being coupled back to $|A\>$ by the 
interaction). Thus the Hamiltonian is $H = H_0 + \varepsilon H^\prime$,
where:

\vspace{0.5\baselineskip}
\noindent (i) $|\psi_0\>$ is an eigenstate of $H_0$;
\vspace{0.5\baselineskip}

\noindent (ii) $H_0$ acts inside $\S^\prime$ to govern the dispersal of
the decay products, so that $e^{-iH_0t}|\psi_1\>$ rapidly becomes
orthogonal to $|\psi_1\>$, and remains so:
\begin{equation}\label{dispersal} 
\<\psi_1|e^{-iH_0t/\hbar}|\psi_1\> = 0 \text{  unless  } t < \tau;
\end{equation}

\noindent (iii)
\upline
\[ 
H^\prime = |BC(0)\>\<A| + |A\>\<BC(0)|.
\]

With this Hamiltonian, the solution of the Schr\"odinger equation with
initial condition $|\psi(t)\> = |\psi_0\>$ at $t=0$ is \cite{obsdecay}

\begin{equation}\label{decay2}
|\psi(t)\> = f(t)|\psi_0\> + \frac{\varepsilon}{i\hbar}\int_0^t
f(t^\prime)e^{-iH_0(t-t^\prime)/\hbar}|\psi_1\>dt^\prime
\end{equation}
where the survival amplitude $f(t)$ is the solution of the
integro-differential equation
\begin{equation}\label{int}
\frac{d}{dt}\left[e^{-iE_0t/\hbar}f(t)\right] =
-\frac{\varepsilon^2}{\hbar^2}
\int_0^t\chi(t-t^\prime)f(t^\prime)e^{iE_0t^\prime/\hbar}dt^\prime 
\end{equation}
with\upline
\[
\chi(t) = \<\psi_1|e^{-iH_0t/\hbar}|\psi_1\>.
\]
According to \eqref{dispersal} the function $\chi(t)$ vanishes for
$|t|<\tau$, so an approximate solution to \eqref{int}, which neglects
the variation of $f$ on time scales of the order of $\tau$, is
\[
f(t) \approx e^{-\Gamma\tau}e^{-iE_0t/\hbar}
\]
with \upline
\[
\Gamma = \frac{\varepsilon^2}{\hbar^2}\int_0^\tau\chi(t)dt.
\]

Now let us adopt Bell's interpretation with just two viable subspaces
$\S_0$ and $\S_1$, where $\S_0$ is the one-dimensional subspace 
spanned by $|\psi_0\>$ and $\S^\prime$ as above is the subspace
containing the states of the decay products. This is to assume that at
any time the system consists either of the unstable particle $A$ or of a
decayed state $B+C$, and it makes transitions between the two with
probabilities given by \eqref{trans1} and \eqref{trans2}. We would expect the
transitions to go only in the direction $A\rightarrow B+C$. 

Using the suffix $d$ (= decayed) to refer to $\S^\prime$, we have
transition probabilities $T_{0d}$ and $T_{d0}$ determined by the real
part of the matrix element 
\[
\frac{1}{i\hbar}\<\psi(t)|\Pi_0 H \Pi_d|\psi(t)\> =
-\frac{\varepsilon^2}{\hbar^2}\overline{f(t)}\int_0^t
q(t-t^\prime)f(t^\prime)dt^\prime.
\]
At times $t\gg\tau$, this is approximately $-\Gamma|f(t)|^2$. Hence the transition
probabilities have exactly their expected values: $T_{0d}=0$ and
$T_{d0}=\Gamma$.

There is still a possibly counter-intuitive element in this picture. If
a transition occurs from the unstable state $|\psi_0\>$ at time $t$, it
goes to the state 
\[
\Pi_d|\psi(t)\> = N^{-1}\int_0^t
f(t^\prime)e^{-iH_0(t-t^\prime)/\hbar}|\psi_1\>dt^\prime
\]
(where $N$ is a normalisation factor) which is a superposition of states
which have decayed at earlier times $t^\prime < t$. If this
superposition is examined to determine the time $t^\prime$ of decay,
there appears to be a possibility that the result of the measurement
will not be the time $t$ at which the transition occurred according to
the stochastic theory, in which case the theory would not be giving an
empirically verifiable account. 

To investigate this, let us now consider a model with a more refined
description of the visible states. If the decayed states in $\S^\prime$
can be examined to determine the time of decay to $|\psi_1\>$, then
$e^{-iH_0t/\hbar}|\psi_1\>$ evolves through a sequence of eigenstates
$|\psi_n\>$ of an observable $T =$ ``time since decay". Suppose that
the eigenvalues of $T$ are spaced at equal intervals $\tau$; then
\eqref{dispersal} is generalised to 
\begin{equation}
\<\psi_n|e^{-iH_0t/\hbar}|\psi_m\> = 0 \text{  unless  } |t - (n-m)\tau|<\tau. 
\end{equation}
Now the transition probability from $|\psi_0\>$ to $|\psi_m\>$ is the
imaginary part of 
\[ 
\<\psi(t)|\Pi_m H \Pi_0|\psi(t)\> = \varepsilon
f(t)\<\psi(t)|\Pi_m|\psi_1\>
\]
which vanishes unless $m=1$. Thus the decay can only be to the correct
state $|\psi_1\>$. This confirms that, if the viable
subspaces are chosen as above, Bell's transition probabilities
give a satisfactory description of the decay process.

\medskip
\begin{center}CHOICES, CHOICES\ldots \end{center}
\smallskip

That last sentence would be a triumphant conclusion if it were not for
the fatal weakness of its qualifying clause. Bell's formulation of
quantum mechanics does not exist until one has specified the viable
subspaces. In this respect it is no improvement on the conventional
formulation, which does not exist until one has specified precisely what
physical arrangements constitute ``measurements". The simplest approach
to the problem in the case of Bell's formulation is to specify a
preferred set of observables -- sorry, beables -- whose eigenspaces will
be the viable subspaces. However, this is bound to seem arbitrary; if
one is aspiring to give an absolute description of the physical world, there
seems to be no good empirical or theoretical reason why any
particular set of physical quantities should have fundamental status.
Bell's own proposal in \cite{Bell:beables} was that these quantities
should be the bilinear invariants of the fundamental fermion fields, but
this only emphasises the arbitrariness: to pick out a particular set of
fermion fields as ``fundamental" abolishes the freedom to make field
redefinitions, which one would like to regard as yielding equivalent
theories. 

A related problem is that of relativistic invariance. A theory based on
transitions between states of the world requires these states to be
defined universally at each instant of time, and therefore has a
definition of simultaneity built in; in other words, it has a preferred 
frame of reference. Since the theory makes the same predictions as
conventional relativistic quantum field theory, this frame of reference
would not be experimentally distinguishable from any other frame related to it
by a Lorentz transformation, but as Bell conceded in the final sentence of
\cite{Bell:beables}, ``It seems an eccentric way to make a world."

There is a way of avoiding the arbitrariness of a choice of preferred
subspaces if the world can be divided into two subsystems, so that the
pilot state space has the structure of a tensor product $\S_1\ox\S_2$. 
In that case any vector in the tensor product canonically defines a
decomposition of the space into orthogonal subspaces by means of the
Schmidt decomposition: given $|\Psi(t)\>\in \S_1\ox\S_2$, we have
\[
\S_1\ox\S_2 = \sum_\lambda \S_{1\lambda}(t)\ox\S_{2\lambda}(t) \oplus
\S^\prime
\]
where $\S_{1\lambda}(t)\subset\S_1$ is the eigenspace of the marginal
density matrix $\rho_1(t)=\tr_2|\Psi(t)\>\<\Psi(t)|$ with eigenvalue
$\lambda$, $\S_{2\lambda}(t)\subset\S_2$ is the eigenspace of
$\rho_2(t)$ with the same eigenvalue, and $\S^\prime$ is the residual
subspace of $\S_1\ox\S_2$, orthogonal to all the
$\S_{1\lambda}\ox\S_{2\lambda}$. Taking the viable subspaces at time $t$ 
to be the $\S_{1\lambda}(t)\ox\S_{2\lambda}(t)$ and $\S^\prime$ gives a 
formulation in which the
subsystem $S_1$ is always in an eigenspace of its density matrix, and
the rest of the world is in a corresponding state.

Such formulations of quantum mechanics, based on the Schmidt
decomposition in a given tensor product structure, are a
subclass of ``modal interpretations" \cite{vanFraassen,
Dieks, BacciaDickson}. They offer the prospect of a theory which
describes real events, reflecting our actual experience of the world, 
without compromising the unitary symmetry of quantum mechanics. 
Unfortunately, the Schmidt decomposition has not
proved true to its promise that it would deliver a canonically defined
set of visible states. It may be granted that the world consists of
subsystems and therefore its state space has a tensor product structure
(though it might also be queried whether this decomposition is uniquely
given as part of nature),
but it is quite clear that there are more than two subsystems. It is
only in a tensor product of two factors, however, that vectors are
guaranteed to have a Schmidt decomposition of the kind used in modal
interpretations; in an $n$-fold tensor product $\S_1\ox\cdots\ox\S_n$ it
is rather unusual \cite{Peres:Schmidt} for vectors to be of the form
\[
\sum_i^N c_i|\psi^{(1)}_i\>\cdots|\psi^{(n)}_i\>
\]
where $|\psi^{(r)}_i\> (i=1,\ldots,N)$ are orthonormal vectors in $\S_r$.
(There is a generalisation of the Schmidt decomposition which is valid
for any vector in an $n$-fold tensor product \cite{Schmidt}, but it does
not relate to the marginal states of the subsystems as the bipartite
Schmidt decomposition does.) In order to take advantage of the properties of
the Schmidt decomposition, therefore, the theory must specify the way
that it divides the world into one particular subsystem and the rest of
the world. We find ourselves once again having to make arbitrary
choices. Desperate attempts to have our cake and eat it are doomed to
failure \cite{Vermaas:nogo}.

\medskip
\begin{center}\ldots AND HOW TO LIVE WITH THEM\end{center}
\smallskip

If arbitrary choices cannot be avoided, we must consider how to live
with them. But we are used to living with arbitrary choices in physics;
they are characteristic of relativistic theories. There is nothing
objectionable about being forced to make an arbitrary choice as long as
we recognise that our descriptions are made relative to that choice,
that any other choice would be equally valid, and as long as we know how to
change from one such choice to another. 

We have just been considering arbitrary choices of preferred
subspaces, preferred physical quantities or preferred decompositions of the world
into subsystems. They seem objectionable if they are seen to form part
of an objective or absolute description of the world, seen from outside.
But there can be no objection to preferred subspaces, observables or
subsystem occurring in a description of the world from inside the world.
Then, indeed, one is obliged to say from where one is observing, what
subsystem of the world is doing the observing, and what observables are
being observed.

This brings us to Everitt's relative-state formulation of quantum
mechanics, which is closely related to Bell's beable formulation and to
modal interpretations, though in some respects it is directly opposed to
them. I would now like to present the relative-state formulation in a 
version which emphasises these similarities and differences, while also
relating it to the conventional interpretation. I hope this will support
a claim that it is the natural context in which to introduce the quantum
jumps whose necessity is the subject of this paper.

\medskip
\begin{center}INTERNAL AND EXTERNAL\end{center}
\smallskip

In Bell's beable interpretation the identities of both the pilot state 
and the visible state are, at any given time, facts about the world.
They have the same propositional status. In the modal interpretation an
identification of the pilot state is a modal statement: it tells us what
can, or must, or might be the case. What \emph{is} the case is the 
identity of the visible state. In Everett's relative-state
interpretation the identities of both the pilot state and the visible
state are facts, but they are facts of different kinds. The difference
can be described by the pairs of words ``absolute" and ``relative", or
by ``objective" and ``subjective". I prefer to use ``external" and
``internal". Most vividly, the distinction has been described by Nagel
\cite{Nagel:nowhere} as that between ``the view from nowhere" and ``the
view from now here".

As always, the best system to illustrate the principle is
Schr\"odinger's cat, with an unstable atom, a diabolical device and a
cat Jacko in a locked box made of unbreakable glass so that the human observer
is helplessly aware of the moment of the cat's death. If the atom is in
its excited state at time $t=0$, the state of our little universe at
time $t$ is given by a vector in the tensor product
$\S_{\text{atom}}\ox\S_{\text{cat}}\ox\S_{\text{human}}$ of the form
\begin{align}\label{cat}
|\Psi(t)\> &= \e^{-\gamma t}
|\text{excited}\>|\text{alive}\>|\text{``Jacko is alive"}\>\notag\\
&+\int_0^t \e^{-\gamma t^\prime}|\text{ground}\>|\text{dead for time }t-t^\prime\>
|\text{``Jacko died at }t^\prime\text{"}\>dt^\prime.
\end{align}
The statements inside the ket signs for the vectors in $\S_{\text{human}}$
indicate brain states in which those statements occur as beliefs. They
are physical configurations; but they are also statements. Are they true
or false? Each is believed by a brain which has observed the fact it
describes, and that fact belongs to reality. As a
human belief, each statement could not be more true. Yet they cannot both
be true, for they contradict each other. I take this to be
characteristic of \emph{internal} statements in a physical system; the
belief of such a statement is a physical occurrence, and its truth can
only be assessed in the physical context in which it occurs. In the
present situation, such a context consists of a particular component of
the universal state \eqref{cat}. 

Internal statements are necessarily part of the experience of a conscious
physical being (I use the word ``conscious" reluctantly, 
because I do not want to be understood as restricting the discussion to
human beings, but it seems to be what I mean). But they are not (or not
solely) \emph{about} experiences -- on
the contrary, they are about the physical world, most interestingly the
part of the world which is separate from the experiencing being. What
makes them internal statements is that they are themselves events in the
physical world. 

The distinction between internal and external statements offers a pat
way to resolve the tension between the two kinds of time development in
the conventional textbook formulation of quantum mechanics: the
continuous, deterministic evolution of the state vector given by the
Schr\"odinger equation, and the discontinuous, probabilistic change
following a measurement, given by the projection postulate. The idea,
roughly speaking, would be that the Schr\"odinger equation is an
external statement, while the projection postulate is an internal one.
More precisely, let $|\Psi(t)\>$ be a time-labelled sequence of states
satisfying the Schr\"odinger equation. Then it is an external (and
tenseless) statement that the world passes through the sequence of
states $|\Psi (t)\>$. Suppose these states are products
$|\phi(t)\>|\psi(t)\>$ where $|\phi(t)\>$ is a state of a measuring apparatus and
a conscious observer, while $|\psi(t)\>$ is the state of the rest of the
world (or simply the system being measured by the apparatus), and
suppose the Hamiltonian includes a measurement (an instantaneous one, since I
am --- for the moment --- in that convention) made at time $t_0$. Then
it is an internal statement (and a tensed one, with a ``now" of $t_0 +
\epsilon$) that after the measurement the state $\phi(t)\>$ has jumped
to an eigenstate of the measured quantity. A report of a measurement
result is an internal statement, made by a conscious system; a general
statement about measurement results must be made relative to a
particular conscious system (just as statements about energy in special
relativity are necessarily relative to a particular frame of reference).

If the argument of this paper is accepted, statements about results
of measurements should be replaced by statements about spontaneous
transitions, or quantum jumps. But these should still be statements of the
same type: a report of a quantum jump is therefore an internal
statement, made by a conscious system, and general statements about
quantum jumps should be made relative to a particular conscious system. 

There is now no problem of arbitrariness in specifying the viable
subspaces between which quantum jumps occur. The second law of quantum
dynamics, governing the transitions of the visible state, is now to be
understood as describing the experiences of conscious systems
--- it is a collection of different laws, one for each conscious system;
and there is no loss of unitary symmetry in the fact that a preferred
system of subspaces occurs in each of these laws. There are no
preferences in the general statement from which the individual
statements of experienced transitions are derived.

There is, however, a preference to be declared for one of two
preference-free formulations of the general law governing experienced
transitions. In describing the experience of a particular conscious
system, the viable subspaces could be determined by means of a
particularly relevant set of observables; or they could be the subspaces
arising in the Schmidt decomposition relative to a particularly relevant
tensor-product split, namely the split into the conscious system and the
rest of the world. At first sight the latter possibility looks more
attractive, perhaps because of a lingering prejudice that Schmidt
decompositions are more canonical than choices of bases. But if we
are avowedly describing the experiences of a stated conscious system,
this already canonically picks out the basis of states of that system in
which it has definite experiences --- what Lockwood \cite{Lockwood:manyminds}
calls the \emph{consciousness basis}, though I will use the term
\emph{experience basis}.\footnote{In passing, let us note that an answer to the
question ``Why don't we see superpositions of macroscopic states?" is
that there is no experience state describing such seeing. A
superposition of two experience states is not an experience
state.} Moreover, if we were to use the Schmidt decomposition there
would be no guarantee that the transitions which the conscious system
experiences would actually take it to states of definite experience. It
is probably true to a high degree of approximation that the states of
the conscious system occurring in the Schmidt decomposition, the
eigenstates of its density matrix, will in fact be eigenstates of
experience; but this will remain an approximation, and if our aim is to
describe experience we should make sure that we actually do so.

We thus arrive at the following general formulation of the laws of
motion in quantum mechanics:

\begin{enumerate}
\item The universe is described by a time-dependent state vector 
$|\Psi(t)\>$ in the universal state space $\S$,
which evolves according to the Schr\"odinger equation. 
\item The experience of any conscious subsystem $C$ of the universe is 
described at any time $t$ by a state vector $|\phi_n\>$ in the experience basis
of that subsystem's state space $\S_C$. If this experience is described by
$|\phi_n\>$ at time $t$, then the probability that it is described by
$|\phi_m\>$ at time $t+\delta t$ is $T_{mn}\delta t$ where
\begin{align}
T_{mn} &= \frac{\max(J_{mn},0)}{\<\psi_n(t)|\psi_n(t)\>},\\
J_{mn} &= 
2\Im\left[\hbar^{-1}\(\<\phi_m|\<\psi_m(t)|\)H\(|\phi_n\>|\psi_n(t)\>\)\right] 
\end{align}
and the states $|\psi_n(t)\>$ are the states of the rest of the world,
(elements of $\S_R$ where $\S = \S_C\ox\S_R$),
which are the coefficients of the experience basis states in the
expansion of the universal state vector with respect to this basis:
\[
|\Psi(t)\> = \sum_n|\phi_n\>|\psi_n(t)\>.
\]
\end{enumerate}

\medskip
\begin{center}
  IS YOUR POSTULATE REALLY NECESSARY?
\end{center}
\smallskip

In the actual practice of physicists giving theoretical analyses of
experiments in quantum optics, like the quantum-jump and watched-pot
experiments described above, it is quite common to base the analysis on
an assumption of quantum jumps --- see, for example, \cite{Carmichael,
Dalibard:MonteCarlo, Hegerfeldt, PlenioKnight, Almut:qjumps, WisemanToombes}. This is
sometimes regarded as a practically convenient method of solving the
equations of motion for a density matrix, sometimes as a description of
a fundamental physical process. If the latter, there is a need for a
basic principle to decide exactly what jumps to what; at present
\cite{WisemanToombes} this decision is an \emph{ad hoc} one, and
different theorists make different assumptions. Attempts to derive a
quantum-jump approach from first principles \cite{Hegerfeldt,
PlenioKnight}, using conventional notions of measurement in quantum
mechanics, fall foul of the watched-pot effect; it becomes necessary to
restrict the application of the measurement postulates in a way which is
once again \emph{ad hoc}. The postulates presented in the previous
section are intended to provide a firm basis from which, in
principle, definite statements about the occurrence of quantum jumps
could be derived.

However, there is another possible attitude. The quantum-jump postulate 
proposed above is a replacement for the conventional measurement
postulate; both function as interpretations of the time-dependent state
vector obtained by solving the Schr\"odinger equation. This is regarded
as a set of different state vectors, one for each time, and each of
these is interpreted; thus there is a set of interpretations (one for
each time in the postulate proposed here; one for each measurement in
the conventional formulation). But one could instead regard the
time-dependent state vector as a single object, a vector-valued function
of a real variable $t$, and attempt to give it a single interpretation.
Time-dependent phenomena like quantum jumps should then emerge by
examining this single interpretation of a time-dependent object.
An experiment extended over time should be regarded as a single
experiment, in which time-dependent phenomena are registered as a record
which can be examined at a later time; a conscious experience extended
over time can be treated as a single experience at a later time which
contains memories of experiences at different times.

Let us see how such an approach could be applied to Dehmelt's
quantum-jump experiment. In the appendix the Schr\"odinger equation is
solved for this system, and it is shown that to a good approximation the
solution can be written as  
\begin{align}
|\Psi(t)\> &= |\psi_0\>\left\{\tilde{f}(t)|0\> +
\sum_n|\tilde{\phi}_n(t)\>\right\} +  |\psi_1\>\left\{\tilde{g}(t)|0\> +
\sum_n|\tilde{\theta}_n(t)\>\right\}\notag\\
&+|\psi_2\>\left\{\tilde{h}(t)|0\> + \sum_n|\tilde{\chi}_n(t)\>\right\},
\label{jumpstate}
\end{align} 
where $|\psi_0\>,|\psi_1\>,|\psi_2\>$ are
the three atomic states shown in Fig.2, and $|\tilde{\phi}_n\>$,
$|\tilde{\theta}_n\>$ and $|\tilde{\chi}_n\>$ are states of the
electromagnetic field, each containing $n$ photons, defined recursively
by 
\begin{equation}\label{fieldstates} 
\begin{aligned}
|\tilde{\phi}_n(t)\> &=
-i\alpha\int_0^\infty\tilde{f}(t-t\pr)\e^{-iH_{\tf}(t-t\pr)}
\left\{a_{\tR}^\dagger|\tilde{\theta}_{n-1}(t\pr)\> +
a_{\tB}^\dagger|\tilde{\chi}_{n-1}(t\pr)\>\right\}dt\pr,\\
|\tilde{\theta}_n(t)\> &=
-i\alpha\int_0^\infty\tilde{g}(t-t\pr)\e^{-iH_{\tf}(t-t\pr)}
\left\{a_{\tR}^\dagger|\tilde{\theta}_{n-1}(t\pr)\> +
a_{\tB}^\dagger|\tilde{\chi}_{n-1}(t\pr)\>\right\}\e^{-i\Omega_{\tR}t\pr}dt\pr,\\
|\tilde{\chi}_n(t)\> &=
-i\alpha\int_0^\infty\tilde{h}(t-t\pr)\e^{-iH_{\tf}(t-t\pr)}
\left\{a_{\tR}^\dagger|\tilde{\theta}_{n-1}(t\pr)\> +
a_{\tB}^\dagger|\tilde{\chi}_{n-1}(t\pr)\>\right\}\e^{-i\Omega_{\tB}t\pr}dt\pr.
\end{aligned} 
\end{equation} 
where $H_{\tf}$ is the Hamiltonian
governing the evolution of the electromagnetic field, and
$a_{\tR}^\dagger$ and $a_{\tB}^\dagger$  are creation operators for the
red and blue photons associated with the excited levels of the atom ---
``emitted when the atom jumps from an excited level to the ground
state", we would say if we had any warrant for talking about quantum
jumps. But now, it could be argued, \eqref{jumpstate} itself gives us
such a warrant. It shows the state at time $t$ as a superposition of
states of the atom and the electromagnetic field which can be distinguished by a
suitable measurement at time $t$. Such a measurement could not only determine
only how many red and blue photons are present --- thus identifying a
field state $|\tilde{\chi}_n(t)\>$, say --- but also, by
measuring how far the photons have moved from the atom, measure the
value of $t\pr$ in, for example, $\e^{-iH_{\tf}(t-t\pr)}
a_{\tR}^\dagger|\tilde{\theta}_{n-1}(t\pr)\>$, i.e. the time at
which the last red photon was emitted. Likewise the times of earlier
emissions could be determined. In this way the history of the quantum
jumps which occurred before time $t$ could be determined by a
measurement at time $t$.

Thus the state $\Psi(t)$ is a superposition of states each of which
contains a record of a particular sequence of quantum jumps. The full
state $\Psi(t)$ evolves continuously, but each component describes
discontinuous events, ultimately because of the quantisation of the
electromagnetic field. Explicit postulation of quantum jumps at a
fundamental level can be replaced by a single application of the usual
measurement postulate at a late time $t$.

There is probably no conclusive argument against this position,
and I suspect that many physicists will find it congenial. Nevertheless,
it strikes me as unsatisfactory. It requires an arbitrary choice of the time $t$
at which the all-deciding measurement is to be made; this can be put off
to a vague ``at the end of the day", but it remains an instant in time,
and nothing can be said about the thereafter. Such an interpretation is
an attempt to step outside time and take an eternal view, as is
certainly possible in classical physics; but it appears not to be
allowed by quantum mechanics. If the interpretation of quantum mechanics is going to
be bound by time eventually, it seems to me to be best not to put it off
and to acknowledge the inescapable importance of time in our experience
and our understanding.

\section*{Appendix: Quantum Theory of Quantum Jumps}

In this appendix we give a conventional quantum-mechanical theory of the
quantum-jump experiment \cite{qjumps:expt1, qjumps:expt2, qjumps:expt3},
solving the full Schr\"odinger equation for the system consisting of the
three-level atom and the electromagnetic field. This does not appear to
have been done in previous theoretical treatments of the experiment.
Most authors \cite{Arecchi, Javanainen, Kimble, Nienhuis, Pegg,
Schenzle} provide this by considering the atomic density matrix, including
off-diagonal elements, and using optical Bloch equations. This, however,
requires an assumption that quantum jumps occur in the crucial
spontaneous emission part of the experiment. The same is true of the
treatment \cite{Cohen-T:qjumps} in terms of the state vector of the
atom. A complete treatment of the atom-field system has been given by
Zoller, Marte and Walls \cite{Zoller} who obtain an equation for the
atomic density matrix in which an effective Hamiltonian, incorporating
the effects of quantum jumps in the same way as in the other papers, is
justified by an appeal to Mollow's fully quantum-mechanical treatment of
resonance fluorescence in two-level atoms \cite{Mollow}, but no such
treatment has been given for the three-level atom involved in Dehmelt's
experiment. Barchielli \cite{Barchielli:qjumps} has given a theory of
the experiment in terms of a quantum stochastic equation. He obtains
this as an approximation to the deterministic Schr\"odinger equation of
the atom-field system, using the same approximations as in the
calculation given here. He does not comment on the origin or physical
significance of the stochastic elements in his treatment.

The theory given here may appear to be less than fully
quantum-mech-anical in that the laser beams are described by classical
oscillating fields, but this has been shown by Moller \cite{Mollow} to
be an accurate description of a coherent quantum field in interaction
with an atom with appropriate boundary conditions at $t=-\infty$. In
order to obtain a solution we need to use the rapid-dispersal
approximation, which can be justified from quantum electrodynamics
\cite{rapid-dispersal}, and to neglect rapidly oscillating quantities,
which is a self-consistency requirement, justified retrospectively. 

The atom is illuminated by two light beams, the red
beam tuned near the frequency $\Omega_{\tR} = (E_1 - E_0)/\hbar$
of the strong transition, and the blue beam
tuned near the frequency $\Omega_{\tB} = (E_2 - E_0)/\hbar$ of the
weak transition to the shelf state. We take these both to be coherent
(laser) beams. Let $\Delta_{\tR}$ and $\Delta_{\tB}$ be their detunings, so
that they are described by classsical fields $\bE_{\tR}\cos[(\Omega_{\tR}
+ \Delta_{\tR})t]$ and $\bE_{\tB}\cos[(\Omega_{\tB} + \Delta_{\tB})t]$. The
corresponding contributions to the atomic Hamiltonian are
\[
H_{\tR} = e\br\cdot\bE_{\tR}\cos[(\Omega_{\tR} + \Delta_{\tR})t] \qquad \text{and}
\qquad H_{\tB} = e\br\cdot\bE_{\tB}\cos[(\Omega_{\tB} + \Delta_{\tB})t]
\]
but the effect of neglecting rapidly oscillating terms can be obtained
by replacing these by 
\begin{equation}
H_{\tR} = \lambda \e^{-i(\Omega_{\tR} + \Delta_{\tR})t}|\psi_1\>\<\psi_0| +
        \lambda^*\e^{i(\Omega_{\tR} + \Delta_{\tR})t}|\psi_0\>\<\psi_1|
\end{equation}
\upline
and
\begin{equation}
H_{\tB} = \Lambda \e^{-i(\Omega_{\tB} + \Delta_{\tB})t}|\psi_2\>\<\psi_0| +
        \Lambda^*\e^{i(\Omega_{\tB} + \Delta_{\tB})t}|\psi_0\>\<\psi_2| 
\end{equation}
where $|\psi_0\>$, $|\psi_1\>$ and $|\psi_2\>$ are the three atomic states and
\[
\lambda = e\<\psi_1|\br\cdot\bE_{\tR}|\psi_0\>, \qquad
\Lambda = e\<\psi_2|\br\cdot\bE_{\tB}|\psi_0\>.
\]

The interaction between the atom and the electromagnetic field, which is
responsible for spontaneous emission from the excited states $|\psi_1\>$ and
$|\psi_2\>$, is described by a Hamiltonian of the form \cite{rapid-dispersal}
\begin{equation}
H_{\text{int}} = \alpha\(|\psi_1\>\<\psi_0|a_{\tR} + |\psi_0\>\<\psi_1|a_{\tR}^\dagger\) + 
\beta\(|\psi_ 2\>\<\psi_0|a_{\tB} + |\psi_0\>\<\psi_2|a_{\tB}^\dagger\)
\end{equation}
where $a_{\tR}^\dagger$, $a_{\tB}^\dagger$, $a_{\tR}$ and $a_{\tB}$ are 
single-photon creation and annihilation operators which create and
annihilate photons associated with the $1\rightarrow 0$ and
$2\rightarrow 0$ transitions (we will call these ``red" and ``blue"
photons, although the photon states created by $a_{\tR}^\dagger$ and
$a_{\tB}^\dagger$ do not have definite frequencies \cite{rapid-dispersal}).

The full Hamiltonian for the atom-field system is 
\begin{equation}
H = H_{\ta} + H_{\tf} + H_{\tR} + H_{\tB} + H_{\text{int}}
\end{equation}
where $H_{\ta}$ is the Hamiltonian for the atom alone, which has
eigenstates $|\psi_0\>,|\psi_1\>,|\psi_2\>$ with eigenvalues $E_0, E_1, E_2$; and
$H_{\tf}$ is the Hamiltonian for the field alone. The state of the whole
system at time $t$ can be written
\begin{equation}
\begin{aligned}\label{6}
|\Psi(t)\> &= \e^{-iE_0t/\hbar}|\psi_0\>\left\{ f(t)|0\> 
+ \sum_{n=1}^\infty|\phi_n(t)\>\right\} \\
& + \e^{-iE_1t/\hbar}|\psi_1\>\left\{ g(t)|0\> +
\sum_{n=1}^\infty|\theta_n(t)\>\right\} e^{-i\Delta_{\tR} t}\\
& + \e^{-iE_2t/\hbar}|\psi_2\>\left\{ h(t)|0\> +
\sum_{n=1}^\infty|\chi_n(t)\>\right\}\e^{-i\Delta_{\tB} t}
\end{aligned}
\end{equation}
where $|0\>$ is the vacuum field state and $|\phi_n(t)\>,
|\theta_n(t)\>, |\chi_n(t)\>$ are $n$-photon states. The Schr\"odinger
equation for $|\Psi(t)\>$, with the Hamiltonian (5), gives
\begin{equation}\label{7}
\begin{aligned}
i\frac{df}{dt} &= \lambda^*g + \Lambda^*h,\\
i\frac{dg}{dt} &= \lambda f + \Delta_{\tR} g + \alpha \e^{i\Omega_{\tR}
t}\<0|a_{\tR}|\phi_1(t)\>,\\ 
i\frac{dh}{dt} &= \Lambda f + \Delta_{\tB} h + \beta \e^{i\Omega_{\tB}
t}\<0|a_{\tB}|\phi_1(t)\>;
\end{aligned}
\end{equation}
\begin{equation}\label{8}
\begin{aligned}
i\frac{d}{dt}|\phi_n\> =& H_{\tf}|\phi_n\> + \lambda^*|\theta_n\> +
\Lambda^*|\chi_n\>\\
&+\alpha\e^{-i\Omega_{\tR} t}a_{\tR}^\dagger|\theta_{n-1}\> +
\beta\e^{-i\Omega_{\tB}t}a_{\tB}^\dagger|\chi_{n-1}\>,\\
i\frac{d}{dt}|\theta_n\> =&\(H_{\tf} + \Delta_{\tR}\)|\theta_n\> +
\lambda|\phi_n\> + \alpha\e^{i\Omega_{\tR} t}a_{\tR}|\phi_{n+1}\>,\\
i\frac{d}{dt}|\chi_n\> =& \(H_{\tf} + \Delta_{\tB}\)|\chi_n\> +
\Lambda|\phi_n\> + \beta\e^{i\Delta_{\tB} t}a_{\tB}|\phi_{n+1}\>,
\end{aligned}
\end{equation}
with the understanding that $|\theta_0(t)\> = g(t)|0\>$ and
$|\chi_0(t)\> = h(t)|0\>$.

We can now use the fact that the spontaneously emitted photons quickly
move away from the vicinity of the atom to eliminate the terms in
$|\phi_1\>$ from eqs. \eqref{7} and the terms containing $|\phi_{n+1}\>$
from \eqref{8}. Let $\mathbf{x} = \(\begin{array}{l}x\\y\\z\end{array}\)$
be an eigenvector of the matrix 
$A = \left(\begin{array}{lll}0&\lambda^*&\Lambda^*\\
\lambda&\Delta_{\tR}&0\\\Lambda&0&\Delta_{\tB}\end{array}\right)$ 
with eigenvalue $\xi$; then eqs. \eqref{8} can be solved for the
combination
\begin{equation}
|\Xi_n(t)\> = x^*|\phi_n(t)\> + y^*|\theta_n(t)\> + z^*|\chi_n(t)\>
\end{equation}
as an integral over the interval $0 < t^\prime < t$ of an integrand
containing $a_{\tR}^\dagger|\theta_{n-1}(t^\prime)\>$,
$a_{\tB}^\dagger|\chi_{n+1}(t^\prime)\>$, $a_{\tR}|\phi_{n+1}(t^\prime)\>$ and
$a_{\tB}|\phi_{n+1}(t^\prime)\>$. The last two can be removed by applying
an annihilation operator $a_{\tB}$ or $a_{\tR}$ and using the
rapid-dispersal approximations \cite{rapid-dispersal}
\begin{equation}
\int_0^t a_X\e^{-iH_{\tf}(t-t^\prime)/\hbar}a_Y|\phi_{n+1}(t^\prime)\> = 0,
\end{equation}
\begin{equation}
a_X\e^{-iH_{\tf}(t-t^\prime)/\hbar}a_Y^\dagger|\Upsilon(t\pr)\> = 0
\quad\mbox{if}\quad t-t\pr \gg \tau,
\end{equation}
\begin{equation}
\int_0^t a_X \e^{-iH_{\tf}(t-t\pr)/\hbar}a_Y^\dagger |\Upsilon(t\pr)\>
\e^{-i\Omega_Y t\pr}F(t\pr)dt\pr =
Q_{XY}|\Upsilon(t)\>F(t)\e^{-i\Omega_Y t}
\end{equation}
where $X$ and $Y$ stand for R or B, $Q_{XY}$ is a constant, $\tau$ is
the dispersal time which is of the order of $10^{-18}$s,
$|\Upsilon(t)\>$ is any of the $n$-photon states of interest, and $F(t)$
is any slowly varying function. This yields
\begin{equation}
a_{\tR}|\Xi_n(y)\> = -i\alpha x^*
\(Q_{\tR\tR}|\theta_{n-1}(t)\>\e^{-i\Omega_{\tR}t} +
Q_{\tR\tB}|\chi_{n-1}(t)\>\e^{-i\Omega_{\tB}t}\)
\end{equation}
and a similar equation with $a_{\tB}$. Note that the right-hand side
does not contain the eigenvalue $\xi$. We can repeat this for each of
the eigenvectors $\mathbf{x}_1, \mathbf{x}_2, \mathbf{x}_3$ of the
matrix $A$; then, using the fact that these eigenvectors satisfy
\[
\mathbf{x}_1\mathbf{x}_1^\dagger + \mathbf{x}_2\mathbf{x}_2^\dagger
+ \mathbf{x}_3\mathbf{x}_3^\dagger = \mathbf{1},
\]
we obtain
\begin{equation}\label{14}
a_{\tR}|\phi_{n+1}(t)\> =
-i\alpha\(Q_{\tR\tR}|\chi_n(t)\>\e^{-i\Omega_{\tR}t} +
Q_{\tR\tB}|\chi_n(t)\>\e^{-i\Omega_{\tB}t}\).
\end{equation}
In particular, for $n=0$ we obtain expressions for
$\<0|a_{\tR}|\phi_1\>$ and $\<0|a_{\tB}|\phi_1\>$ in terms of $g(t)$ and
$h(t)$. Putting these into eq. \eqref{7} and ignoring rapidly
oscillating terms with the frequency $\Omega_{\tB} -\Omega_{\tR}$ gives
a set of homogeneous equations for the amplitudes $f,g,h$ which are the
same as those obtained by intuitive arguments assuming quantum jumps:
\begin{equation}\label{15}
\begin{aligned}
\frac{df}{dt} &= -i\lambda^*g-i\Lambda^*h,\\
\frac{dg}{dt} &= -i\lambda f - (\gamma_{\tR} + i\Delta_{\tR})g,\\
\frac{dh}{dt} &= -i\Lambda f - (\gamma_{\tB} + i\Delta_{\tB})h
\end{aligned}
\end{equation}
where
\[
\gamma_{\tR} = \frac{\alpha^2Q_{\tR\tR}}{\hbar^2} \qquad \mbox{and}
\qquad \gamma_{\tB} = \frac{\alpha^2 Q_{\tB\tB}}{\hbar^2}
\]
are the (complex) decay rates (including energy shifts) for the
spontaneous transitions $1\rightarrow 0$ and $2\rightarrow 0$.

In general, using \eqref{14} to eliminate $|\phi_{n+1}\>$ from the
equations \eqref{8} for the $n$-photon states
$|\phi_n\>,|\theta_n\>,|\chi_n\>$ yields inhomogeneous equations which
have the same matrix as \eqref{15} in their homogeneous part, with a
source term for $|\phi_n\>$ involving the $(n-1)$-photon states
$|\theta_{n-1}\>$ and $|\chi_{n-1}\>$:
\begin{equation}\label{16}
\begin{aligned}
i\frac{d}{dt}|\phi_n\> =& H_{\tf}|\phi_n\> + \lambda^*|\theta_n\>
+\Lambda^*|\chi_n\>\\
&+\alpha\e^{-i\Omega_{\tR}t}a_{\tR}^\dagger|\theta_{n-1}\> +
\alpha\e^{-i\Omega_{\tB}t}a_{\tB}^\dagger|\chi_{n-1}\>,\\
i\frac{d}{dt}|\theta_n\> =& \lambda|\phi_n\> + 
(H_{\tf} + \Delta_{\tR} -i\gamma_{\tR})|\theta_n\>\\
i\frac{d}{dt}|\chi_n\> =& \Lambda|\phi_n\> + (H_{\tf} + \Delta_{tB}
-i\gamma_{tB})|\chi_n\>
\end{aligned}
\end{equation}
(after dropping rapidly oscillating terms).

Suppose the atom is in its ground state at $t=0$, no photons having been
emitted; then the initial conditions are $f(0)=1$, $g(0)=h(0)=0$ and
$|\phi_n(0)\> = |\theta_n(0)\> = |\chi_n\> = 0$. The solution
$(f(t),g(t),h(t))$ of the homogeneous equations \eqref{15} with these
initial conditions can then be used as a Green's function for the
inhomogeneous equations \eqref{16} (after taking account of the field
Hamiltonian $H_{\tf}$ by multiplying by $\e^{-iH_{\tf}t}$). The result
is
\begin{equation}
\begin{aligned}\label{17}
|\phi_n(t)\> =& -i\alpha\int_0^t\e^{-iH_{\tf}(t-t\pr)}f(t-t\pr)\\
&\left\{a_{\tR}^\dagger|\theta_{n-1}(t\pr)\>\e^{-i\Omega_{\tR}t\pr}
+ a_{\tB}^\dagger|\chi_{n-1}(t\pr)\e^{-i\Omega_{\tB}t\pr}\right\}dt\pr
\end{aligned}
\end{equation}
with similar expressions for $|\theta_n(t)\>$ and $|\chi_n(t)\>$. It is
more meaningful to present these in terms of the actual $n$-photon
states, without the exponential factors that were introduced into the
definition \eqref{6} in order to simplify the calculations. Thus, if we
define quantities with a tilde so that the full state at time $t$ is 
\begin{equation}
\begin{aligned}
|\Psi(t)\> &=
|\psi_0\>\left\{\tilde{f}(t)|0\> + \sum_n|\tilde{\phi}_n(t)\>\right\} + 
|\psi_1\>\left\{\tilde{g}(t)|0\> + \sum_n|\tilde{\theta}_n(t)\>\right\}\\
+&|\psi_2\>\left\{\tilde{h}(t)|0\> + \sum_n|\tilde{\chi}_n(t)\>\right\},
\end{aligned}
\end{equation}
then the $n$-photon states $|\tilde{\phi}_n\>$, $|\tilde{\theta}_n\>$ and
$|\tilde{\chi}_n\>$ are given by
\begin{equation}
\begin{aligned}\label{19}
|\tilde{\phi}_n(t)\> &=
-i\alpha\int_0^\infty\tilde{f}(t-t\pr)\e^{-iH_{\tf}(t-t\pr)}
\left\{a_{\tR}^\dagger|\tilde{\theta}_{n-1}(t\pr)\> +
a_{\tB}^\dagger|\tilde{\chi}_{n-1}(t\pr)\>\right\}dt\pr,\\
|\tilde{\theta}_n(t)\> &=
-i\alpha\int_0^\infty\tilde{g}(t-t\pr)\e^{-iH_{\tf}(t-t\pr)}
\left\{a_{\tR}^\dagger|\tilde{\theta}_{n-1}(t\pr)\> +
a_{\tB}^\dagger|\tilde{\chi}_{n-1}(t\pr)\>\right\}\e^{-i\Omega_{\tR}t\pr}dt\pr,\\
|\tilde{\chi}_n(t)\> &=
-i\alpha\int_0^\infty\tilde{h}(t-t\pr)\e^{-iH_{\tf}(t-t\pr)}
\left\{a_{\tR}^\dagger|\tilde{\theta}_{n-1}(t\pr)\> +
a_{\tB}^\dagger|\tilde{\chi}_{n-1}(t\pr)\>\right\}\e^{-i\Omega_{\tB}t\pr}dt\pr.
\end{aligned}
\end{equation}

As in the case of resonance fluorescence in two-state atoms
\cite{rapid-dispersal}, these expressions have a simple intuitive
analysis as superpositions of states in which the atom was in an excited
state at some time $t\pr$, having previously emitted $n-1$ photons,
emitted the $n\/$th photon at time $t\pr$, and emitted no further
photons between times $t\pr$ and $t$. (This description, however, does
not account for the exponential factors $\e^{-i\Omega_{\tR}t\pr}$ and
$\e^{-i\Omega_{\tB}t\pr}$ in $|\tilde{\theta}_n(t)\>$ and
$|\tilde{\chi}_n(t)\>$.) The rapid-dispersal property of the photons
created by $a_{\tB}^\dagger$ and $a_{\tR}^\dagger$ ensures that the
different states in the superpositions do not interfere with each other,
so that the probabilities for the existence of $n$ photons follow
classical laws; for example, the probability that there are $n$ photons
at time $t$ and the atom is in its red state is
\begin{equation}
\begin{aligned}
\<\tilde{\theta}_n(t)|\tilde{\theta}_n(t)\>& = 2\Re\gamma_{\tR}
\int_0^t|\tilde{g}(t-t\pr)|^2
\<\tilde{\theta}_{n-1}(t\pr)|\tilde{\theta}_{n-1}(t\pr)\>dt\pr\\
&+2\Re\gamma_{\tB}\int_0^t|\tilde{g}(t-t\pr)|^2
\<\tilde{\chi}_{n-1}(t\pr)|\tilde{\chi}_{n-1}(t\pr)\>dt\pr 
\end{aligned}
\end{equation}
which is the probability that would be obtained by a naive argument
assuming that the atom is always in one of its three states, that when
it is in one of the two higher states it has a probability
$(2\Re\gamma)dt$ of decaying into the ground state in time $dt$, and that
if it is in the ground state at time $t\pr$ the probability that it will
be in the first excited state at a later time $t$ is
$|\tilde{g}(t-t\pr)|^2$. 

Eqs. \eqref{19} contain the full paradox of the description of
time development in quantum mechanics. They represent the development of
the system as being continuous in time, and yet they describe an
experience (that of watching the trapped atom) which is undeniably
discontinuous. This emphasises the shadowy status of the state vector,
whose probabilistic and ontological aspects coexist uneasily. Insofar as
the state vector has the nature of a probability, it is not surprising
that it should change continuously while describing a discontinuous
process; this is normal for a discrete stochastic process. In such a
process, however, the states are described quite separately from the
probabilities. The formulation of quantum mechanics suggested in the
body of the paper can be seen as restoring this separation in the 
distinction between the pilot state and the visible state, the former
being part of the statement of probabilities while the latter
is (in classical terms) a state specification.


\end{document}